\documentclass[prd,twocolumn,eqsecnum,floats,nofootinbib]{revtex4}
\usepackage{bm}
\usepackage{amssymb}
\usepackage{graphicx}
\begin{document}
\title{Mass change and motion of a scalar charge in cosmological
       spacetimes}  
\author{Roland Haas and Eric Poisson}
\affiliation{Department of Physics, University of Guelph, Guelph,
Ontario, Canada N1G 2W1}
\date{November 22, 2004} 
\begin{abstract}
Continuing previous work reported in an earlier paper [L.M.\ Burko,
A.I.\ Harte, and E.\ Poisson, {\it Phys. Rev. D} {\bf 65}, 124006
(2002)] we calculate the self-force acting on a point scalar charge  
in a wide class of cosmological spacetimes. The self-force produces
two types of effect. The first is a time-changing inertial mass, and
this is calculated exactly for a particle at rest relative to the
cosmological fluid. We show that for certain cosmological models,  
the mass decreases and then increases back to its original value. For 
all other models except de Sitter spacetime, the mass is restored only
to a fraction of its original value. For de Sitter spacetime the mass
steadily decreases. The second effect is a deviation relative to
geodesic motion, and we calculate this for a charge that moves slowly
relative to the dust in a matter-dominated cosmology. We show that the
net effect of the self-force is to push on the particle. We show that
this is not an artifact of the scalar theory: The electromagnetic
self-force acting on an electrically charged particle also pushes on
the particle. The paper concludes with a demonstration that the
pushing effect can also occur in the context of slow-motion
electrodynamics in flat spacetime.  
\end{abstract}
\maketitle

\section{Introduction} 
\label{i}

The prospect of a near-future measurement of gravitational waves
produced by the capture of a solar-mass compact object by a massive
black hole (one of the identified sources for the Laser Interferometer
Space Antenna \cite{LISA}) is motivating an intense effort to
calculate the motion of a compact body in a small-mass-ratio
approximation that goes beyond the test-mass description. In this
approximation the gravitational influence of the smaller body is taken
into account, but the motion is still described as taking place in the
background spacetime of the larger body. While the motion is geodesic
in the test-mass description, it is accelerated in the
small-mass-ratio approximation, and the small body is said to move
under the influence of its gravitational self-force. The work that has
gone into formulating and evaluating the gravitational self-force is
reviewed in detail in Refs.~\cite{poisson:04b, poisson:04c}, and a
shorter, less technical account is given in Ref.~\cite{poisson:04d}. 

The calculation of the gravitational self-force involves a number of
technical and conceptual difficulties, and concrete results pertaining
to situations of astrophysical interest have not yet been obtained. 
To shed light on the physics of self-forces in curved spacetime it is
helpful to consider toy problems. In this paper we continue an
effort begun in an earlier paper (Ref.~\cite{burko-etal:02}, hereafter
cited as ``paper I'') to study the self-force acting on a point scalar 
charge inserted in a cosmological spacetime. The self-force acting on
a scalar charge displays many similarities with the gravitational
self-force (including a dependence on the particle's past history),
but it is technically much easier to obtain (fewer field components
are involved). And the cosmological models considered here are
sufficiently simple that the scalar self-force problem can be solved
in a countably infinite number of examples. Beyond these technical
advantages, the physics of a scalar point charge is surprisingly
interesting: Because a scalar field can radiate in monopole waves, the
inertial mass of a scalar charge is not necessarily a constant of the
motion. This phenomenon of mass radiation is one of our main topics in
this paper. We also calculate the motion of a scalar charge when it is 
subjected to its self-force, and compare this with the motion of an
electric charge. 

The physics of scalar charges in curved spacetime was first
investigated by Quinn \cite{quinn:00}; his theory is briefly
reviewed in Sec.~\ref{ii}. In paper I \cite{burko-etal:02} Quinn's
theory was applied to a scalar particle at rest in a spatially-flat
cosmological spacetime with metric 
\begin{equation} 
ds^2 = a^2(\eta)\bigl(-d\eta^2 + dx^2 + dy^2 + dz^2 \bigr), 
\label{1.1}
\end{equation} 
where $\eta$ is conformal time and $(x,y,z)$ are comoving spatial
coordinates. In paper I the scale factor was restricted to the forms
$a(\eta) = C \eta^2$ (a matter-dominated cosmology) and $a(\eta) =
C/|\eta|$ (de Sitter spacetime); $C$ is a scaling constant. For the 
matter-dominated cosmology it was shown that the scalar charge
initially loses a fraction (possibly all) of its mass, but that this
radiated mass eventually returns to the particle. For de Sitter
spacetime it was shown that the particle radiates all of its mass
within a finite proper time.  

In this paper we generalize the work reported in paper I
\cite{burko-etal:02} on several fronts. We consider a much larger
class of cosmological models, those for which $a(\eta) =
C|\eta|^\alpha$ with $\alpha$ restricted to (positive and negative)
integers. For this class of models we obtain an exact expression for
$m(\eta)$, the particle's mass function. We show that for all values
of $\alpha$ except $\alpha = -1$ (de Sitter spacetime), the mass first
decreases and then increases back to a fraction of its original value;
when $\alpha$ is positive this fraction is in fact equal to unity.  

We also calculate the self-force acting on a scalar charge that moves
relative to the cosmological fluid. Here our treatment is no longer as
general: it is restricted by a slow-motion assumption and to a
matter-dominated cosmology. We show that the net effect of the
self-force is to push on the particle: When a charged particle and an
uncharged particle start with the same initial velocity, the charged
particle subsequently moves faster than the uncharged particle;
ultimately both particles are brought to rest by the cosmological
expansion. While this conclusion may appear surprising, we show that
it is not an artifact of the scalar theory: the same conclusion
applies to an electrically charged particle.  

We begin in Sec.~\ref{ii} with a description of our cosmological
spacetimes and a review of Quinn's theory of point scalar charges. The
time-changing mass and the motion of the particle can both be
determined once the retarded Green's function for the scalar wave
equation has been computed; this computation is carried out in
Sec.~\ref{iii}. In Sec.~\ref{iv} we examine the inertial mass of a
scalar charge at rest in our cosmological spacetimes, and establish
the behaviors that were described previously. In Sec.~\ref{v} we
calculate the motion of a slowly-moving scalar charge in a
matter-dominated cosmology, and show that the net effect of the
self-force is to push on the particle. In Sec.~\ref{vi} we carry out a
similar calculation for an electrically charged particle, but we
generalize the cosmological models to the class described by 
$a(\eta) = C \eta^\alpha$ for any (real or integer, but positive)
$\alpha$; we show that the electromagnetic self-force also pushes on
the particle. In our final section \ref{vi} we reflect on this perhaps 
surprising property of the self-force, that it accelerates the
particle instead of slowing it down (relative to the motion of an
uncharged particle); this goes against a naive expectation that the
self-force should always give rise to radiative damping. In
Sec.~\ref{vi} we show that this property is not an artifact of curved
spacetime: It can be manifested also in the simplest context of
slow-motion electrodynamics in flat spacetime.     

\section{Equations of motion of a scalar charge in a cosmological
spacetime}  
\label{ii}

A particle of mass $m$ and scalar charge $q$ moves in a cosmological
spacetime with metric 
\begin{equation} 
ds^2 = a^2(\eta) \bigl( -d\eta^2 + dx^2 + dy^2 + dz^2 \bigr), 
\label{2.1}
\end{equation} 
where $\eta$ is conformal time and $\bm{x} = (x,y,z)$ are comoving
spatial coordinates; we use $x \equiv (\eta,\bm{x})$ to label a 
spacetime event, and boldfaced symbols represent three-dimensional
vectors in flat space. The cosmology is restricted to be spatially
flat, and the scale factor is restricted to the form 
\begin{equation}
a(\eta) = C |\eta|^\alpha, 
\label{2.2}
\end{equation}
where $C$ is a scaling constant and $\alpha$ is a constant (positive
or negative) exponent. When $\alpha > 0$ we take $\eta$ to be positive
and increasing from $0$ to $\infty$. When $\alpha < 0$ we take $\eta$
to be negative and increasing from $-\infty$ to $0$. In all cases
$a(\eta)$ increases with increasing $\eta$. The particle moves on a
world line $\gamma$ described by the parametric relations
$z^\mu(\tau)$, with $\tau$ denoting proper time; its velocity vector
is $u^\mu = dz^\mu/d\tau$.    

The mass density of the cosmological fluid is 
\begin{equation}
\rho = \frac{3 \alpha^2}{8\pi C^2} \frac{1}{|\eta|^{2\alpha+2}} 
\label{2.3}
\end{equation}
and its pressure is given by 
\begin{equation} 
p = \frac{2-\alpha}{3\alpha} \rho. 
\label{2.4}
\end{equation} 
Imposing the dominant energy condition (see, for example, Sec.~9.2 of 
Ref.~\cite{wald:84} or Sec.~2.1 of Ref.~\cite{poisson:b04}) restricts 
$\alpha$ to the interval $\alpha \geq \frac{1}{2}$ if it is positive,
and $\alpha \leq -1$ if it is negative. For $\alpha = 0$ we have flat
spacetime, for $\alpha = 1$ a radiation-dominated cosmology ($p =
\frac{1}{3} \rho$), for $\alpha = 2$ a matter-dominated cosmology ($p
= 0$), and for $\alpha = -1$ we have de Sitter spacetime ($p=-\rho$). 

The scalar field $\Phi$ produced by the scalar charge $q$ obeys the
wave equation 
\begin{equation} 
g^{\alpha\beta} \nabla_\alpha \nabla_\beta \Phi = -4\pi \mu, 
\label{2.5}
\end{equation}
in which $\nabla_\alpha$ indicates covariant differentiation and 
\begin{equation} 
\mu(x) = q \int_\gamma \delta_4\bigl(x,z(\tau)\bigr)\, d\tau 
\label{2.6}
\end{equation}
is the scalar charge density; here $\delta_4(x,x')$ is a scalarized,
four-dimensional Dirac distribution that satisfies 
$\int \delta_4(x,x') \sqrt{-g}\, d^4x = 1$ when $x'$ is within the
domain of integration. The physical solution to Eq.~(\ref{2.5}) is
written as  
\begin{equation} 
\Phi(x) = q \int_\gamma G\bigl(x,z(\tau)\bigr)\, d\tau, 
\label{2.7}
\end{equation}
in terms of a retarded Green's function $G(x,x')$ which is a solution 
to 
\begin{equation} 
g^{\alpha\beta} \nabla_\alpha \nabla_\beta G(x,x') = -4\pi
\delta_4(x,x'). 
\label{2.8}
\end{equation}
When viewed as a function of $x'$, the Green's function has support
both on and within the past light cone of the spacetime event $x$; the
contribution from the light cone is singular, but the contribution
from the interior is smooth.    

The field $\Phi(x)$ is singular on $\gamma$ but its action on the
scalar charge is nevertheless well defined. As shown by Quinn
\cite{quinn:00} (and reviewed in Refs.~\cite{poisson:04b} and
\cite{poisson:04c}), the equations of motion of a freely moving scalar
charge are  
\begin{eqnarray} 
m \frac{D u^\mu}{d\tau} &=& q^2 \bigl( \delta^\mu_{\ \nu} 
+ u^\mu u_\nu \bigr) \biggl[ \frac{1}{6} R^\nu_{\ \lambda} 
u^\lambda 
\nonumber \\ & & \mbox{} 
+ \int_{0}^{\tau-\epsilon} \nabla^\nu
G\bigl(z(\tau),z(\tau')\bigr)\, d\tau' \biggr]  
\label{2.9}
\end{eqnarray} 
and 
\begin{equation}
\frac{dm}{d\tau} = -q^2 \biggl[ \frac{1}{12} {\cal R}  
+ u^\mu \int_{0}^{\tau-\epsilon}
\nabla_\mu G\bigl(z(\tau),z(\tau')\bigr)\, d\tau' \biggr]. 
\label{2.10}
\end{equation} 
The left-hand side of Eq.~(\ref{2.9}) is the product of $m(\tau)$
with the particle's covariant acceleration vector, and the right-hand 
side is the scalar self-force. This involves the spacetime's Ricci
tensor and the particle's velocity vector evaluated at the current
position $z(\tau)$ on the world line. It also involves the gradient of 
the retarded Green's function evaluated at $z(\tau)$ and a prior
position $z(\tau')$ --- the Green's function is differentiated with
respect to $x$ before making the identification $x =
z(\tau)$. Finally, the self-force involves an integration over the
past portion of the particle's world line, from the big bang at 
$\tau' = 0$ to almost $\tau' = \tau$; the integration is cut short by
$\epsilon = 0^+$ to avoid the singular behavior of the Green's
function at coincidence. The right-hand side of Eq.~(\ref{2.10}) gives
the rate at which the particle's inertial mass changes with proper
time $\tau$; it involves the same objects as Eq.~(\ref{2.9}), together
with ${\cal R} = R^\mu_{\ \mu}$, the spacetime's Ricci scalar evaluated
at $z(\tau)$.   

To solve Eqs.~(\ref{2.5}) and (\ref{2.8}) in the cosmological
spacetimes it is useful to introduce the reduced field 
\begin{equation}
\psi(\eta,\bm{x}) = a(\eta) \Phi(x),  
\label{2.11}
\end{equation}
the reduced density $\mu^*(\eta,\bm{x}) = a^3(\eta) \mu(x)$, and 
the reduced Green's function 
\begin{equation}
g(\eta,\bm{x};\eta',\bm{x'}) = a(\eta) a(\eta') G(x,x'). 
\label{2.12}
\end{equation} 
These satisfy 
\begin{equation} 
\biggl( -\frac{\partial^2}{\partial \eta^2} + \nabla^2 
+ \frac{\alpha(\alpha-1)}{\eta^2} \biggr) \psi = -4\pi \mu^* 
\label{2.13}
\end{equation}
and 
\begin{equation} 
\biggl( -\frac{\partial^2}{\partial \eta^2} + \nabla^2 
+ \frac{\alpha(\alpha-1)}{\eta^2} \biggr) g = -4\pi \delta(\eta-\eta')
\delta_3(\bm{x} - \bm{x'}).
\label{2.14}
\end{equation}
In these equations $\nabla^2$ stands for the Laplacian
operator in three-dimensional flat space, and
$\delta_3(\bm{x}-\bm{x'}) = \delta(x-x') \delta(y-y') \delta(z-z')$ is
an ordinary three-dimensional Dirac distribution. The reduced density
obtained from Eq.~(\ref{2.6}) is 
\begin{equation} 
\mu^*(\eta,\bm{x}) = \frac{q}{\gamma} 
\delta_3\bigl(\bm{x} - \bm{z}(\eta)\bigr), 
\label{2.15}
\end{equation}
where $\gamma \equiv a(\eta) d\eta/d\tau 
= (1-\bm{v} \cdot \bm{v})^{-1/2}$, with $\bm{v}(\eta) = d\bm{z}/d\eta$
denoting the particle's coordinate velocity vector and $\bm{v} \cdot
\bm{v} \equiv (v^x)^2 + (v^y)^2 + (v^z)^2$. The retarded solution to
Eq.~(\ref{2.13}) is 
\begin{equation} 
\psi(\eta,\bm{x}) = \int g(\eta,\bm{x};\eta',\bm{x'})
\mu^*(\eta',\bm{x'})\, d\eta' d^3x', 
\label{2.16} 
\end{equation} 
and this statement is equivalent to Eq.~(\ref{2.7}). 

\section{Reduced Green's function} 
\label{iii} 

The difficulty in evaluating Eqs.~(\ref{2.9}) and (\ref{2.10}) resides  
mostly in the computation of the retarded Green's function. In paper I 
\cite{burko-etal:02} the Green's function was obtained for two
specific cosmological models, those for which $\alpha = 2$ (matter
dominated) and $\alpha = -1$ (de Sitter). In this section we
generalize this calculation and compute the Green's function for
arbitrary, but integer, values of $\alpha$. 

For convenience we introduce a positive integer $l$ such that 
\begin{equation}
\alpha = -l \qquad \mbox{when $\alpha \leq 0$} 
\label{3.1}
\end{equation}
and 
\begin{equation} 
\alpha = l + 1 \qquad \mbox{when $\alpha > 0$}. 
\label{3.2}
\end{equation}
We notice that with this substitution, $\alpha(\alpha-1) = l(l+1)$. We
notice further that a given value of $l$ produces two distinct
cosmological models, one with $\alpha_1 = -l \leq 0$ and another with 
$\alpha_2 = l + 1 > 0$; these models are ``equivalent'' in the sense
that $\alpha_1 (\alpha_1 - 1) = \alpha_2 (\alpha_2 - 1) = l(l+1)$, so
that they give rise to the same reduced Green's function. In this
sense, flat spacetime ($\alpha = 0$) is equivalent to a
radiation-dominated cosmology ($\alpha = 1$), and de Sitter spacetime
($\alpha = -1$) is equivalent to a matter-dominated cosmology ($\alpha 
= 2$). 

It was explained in some detail in paper I \cite{burko-etal:02} that
for the cosmological spacetimes considered in this paper, the general
structure of the reduced Green's function is 
\begin{eqnarray} 
g(\eta,\bm{x};\eta',\bm{x'}) &=& \frac{\delta(\eta-\eta'-R)}{R} 
\nonumber \\ & & \mbox{} 
+ \theta(\eta-\eta'-R) V(\eta,\bm{x};\eta',\bm{x'}), \qquad
\label{3.3}
\end{eqnarray}
where $R \equiv |\bm{x}-\bm{x'}| \equiv \sqrt{(x-x')^2 + (y-y')^2 
+ (z-z')^2}$ and $V(\eta,\bm{x};\eta',\bm{x'})$ 
is a smooth two-point function that satisfies the homogeneous form of
Eq.~(\ref{2.14}). The first term on the right of Eq.~(\ref{3.3}) is
the singular contribution from the light cone; the second term is the
contribution from the cone's interior. In flat spacetime $V = 0$, and
the support of the Green's function is entirely on the light
cone. Because a radiation-dominated universe ($\alpha=1$) is
equivalent to flat spacetime (in the sense defined in the previous
paragraph), we have that $V = 0$ in this case as well.\footnote{This 
conclusion follows also from the fact that ${\cal R} = 0$ for a 
radiation-dominated cosmology, which implies that the scalar wave
equation is then conformally invariant. Because the spacetime is
conformally related to flat spacetime, the cosmological and Minkowski 
solutions are related simply by a conformal factor.} In the following
we will exclude $l = 0$ from our considerations, and we will compute
the reduced Green's function for $\eta - \eta' > R$; in this domain we
have $g = V$.    

A Fourier-transform method to solve Eq.~(\ref{2.14}) was described in 
Sec.~IV B of paper I \cite{burko-etal:02}, and the reader is referred
to this paper for details. The core of the method consists of
integrating the ordinary differential equation 
\begin{equation} 
\biggl( \frac{d^2}{d\eta^2} + k^2 - \frac{l(l+1)}{\eta^2} \biggr)
\hat{g} = 0
\label{3.4}
\end{equation}
for the function $\hat{g}(\eta,\eta';k)$, subjected to the boundary 
conditions   
\begin{equation} 
\hat{g}(\eta=\eta';k) = 0, \qquad
\frac{d \hat{g}}{d\eta}(\eta=\eta';k) = 4\pi. 
\label{3.5}
\end{equation}
The retarded Green's function is then obtained by inverting the
Fourier transform, which leads to 
\begin{eqnarray}
V &=& \frac{1}{2\pi^2 R} \int_0^\infty 
\hat{g}(\eta,\eta';k) k \sin(k R)\, dk 
\nonumber \\ 
&=& -\frac{1}{2\pi^2 R} \frac{\partial}{\partial R} \int_0^\infty  
\hat{g}(\eta,\eta';k) \cos(k R)\, dk 
\label{3.6}
\end{eqnarray}
when $\eta - \eta' > R$. The second representation will be more useful
for our purposes. 

The general solution to Eq.~(\ref{3.4}) is $\hat{g} = A k\eta
j_l(k|\eta|) + B k\eta n_l(k|\eta|)$, which is expressed in terms of 
spherical Bessel functions. The boundary conditions of
Eqs.~(\ref{3.5}) determine the constants $A$ and $B$, and their
form can be simplified by invoking the Wronskian-type relations that
are satisfied by the Bessel functions (see, for example, Sec.~10.1.6
of Ref.~\cite{abramowitz-stegun:72}). We find $A = -4\pi |\eta'|
n_l(k|\eta'|)$, $B = 4\pi |\eta'| j_l(k|\eta'|)$, and  
\begin{equation} 
\hat{g} = 4\pi k \eta |\eta'| \bigl[ j_l(k|\eta'|) n_l(k|\eta|) 
- n_l(k|\eta'|) j_l(k|\eta|) \bigr]. 
\label{3.7}
\end{equation} 
After substituting this result into Eq.~(\ref{3.6}) and using the
identities (see, for example, Secs.~9.1.2, 10.1.1, and 10.1.12 of  
Ref.~\cite{abramowitz-stegun:72}) $j_l(z) = (\pi/2z)^{1/2}
J_{l+\frac{1}{2}}(z)$, $n_l(z) = (-1)^{l+1} (\pi/2z)^{1/2}
J_{-l-\frac{1}{2}}(z)$, and $\cos(z) = (\pi z/2)^{1/2}
J_{-\frac{1}{2}}(z)$, we obtain
\begin{widetext}  
\begin{equation} 
V = \mbox{sign}(\alpha) (-1)^l \sqrt{ \frac{\pi}{2} } \frac{1}{R} 
\frac{\partial}{\partial R} \sqrt{\eta \eta' R}   
\int_0^\infty \sqrt{k} \Bigl[ J_{l+\frac{1}{2}}(k|\eta'|)
J_{-l-\frac{1}{2}}(k|\eta|) 
- J_{-l-\frac{1}{2}}(k|\eta'|) J_{l+\frac{1}{2}}(k|\eta|) 
\Bigr] J_{-\frac{1}{2}}(kR) \, dk,
\label{3.8}
\end{equation}
in which all functions are Bessel functions of the first kind. The
expression also involves $\mbox{sign}(\alpha) \equiv \mbox{sign}(\eta)
\equiv \eta/|\eta|$; recall that $\eta$ and $\eta'$ are both negative
when $\alpha < 0$.  

The integrals on the right-hand side of Eq.~(\ref{3.8}) can be
evaluated in closed form, thanks to Eq.~(6.578.1) of
Ref.~\cite{gradshteyn-ryzhik:80}. In the next paragraph we shall
describe the calculation of $V$ for $\alpha > 0$ only; a very similar 
calculation that would return $V$ for $\alpha < 0$ will not be
presented.   

The first integral on the right-hand side of Eq.~(\ref{3.8}) is
proportional to $1/\Gamma(-l)$ and it vanishes when $l$ is an
integer. The second integral is equal to 
\[
-\frac{(\eta/\eta')^l}{\Gamma(-l+{\textstyle \frac{1}{2}})
\Gamma(l+1)} \sqrt{\frac{2}{\eta\eta' R}} F_4\bigl[-l,
{\textstyle \frac{1}{2}};-l+{\textstyle \frac{1}{2}},
{\textstyle \frac{1}{2}};(\eta'/\eta)^2,(R/\eta)^2 \bigr], 
\]
\end{widetext} 
and it is expressed in terms of Appell's hypergeometric function of
two variables (see, for example, Sec.~9.18 of
Ref.~\cite{gradshteyn-ryzhik:80}). Because its first argument is a
negative integer, the Appell function is in fact a terminating
polynomial in the two variables $(\eta'/\eta)^2$ and $(R/\eta)^2$.     
After cleaning up the Gamma functions we arrive at 
\begin{eqnarray}
V &=& - \frac{(2l-1)!!}{(2l)!!} \frac{(\eta/\eta')^l}{R} 
\nonumber \\ & & \mbox{} \times 
\frac{\partial}{\partial R} 
F_4\bigl[-l,{\textstyle \frac{1}{2}};-l+{\textstyle \frac{1}{2}},
{\textstyle \frac{1}{2}};(\eta'/\eta)^2,(R/\eta)^2 \bigr] \qquad
\label{3.9}
\end{eqnarray}
for the reduced Green's function. There remains the task of evaluating
the derivative with respect to $R$. We use the identity 
\[
\frac{\partial}{\partial y} F_4(\alpha,\beta;\gamma,\gamma';x,y) =
\frac{\alpha\beta}{\gamma'}
F_4(\alpha+1,\beta+1;\gamma,\gamma'+1;x,y), 
\]
which is easily established from the series representation of the
Appell function. Our final result is 
\begin{eqnarray} 
V(\eta,\bm{x};\eta',\bm{x'}) &=& \frac{(2l-1)!!}{(2l-2)!!}
\frac{(\eta/\eta')^l}{\eta^2} 
\nonumber \\ & & \hspace*{-80pt} \mbox{} \times
F_4\bigl[-l+1,{\textstyle \frac{3}{2}};-l+{\textstyle \frac{1}{2}}, 
{\textstyle \frac{3}{2}};(\eta'/\eta)^2,(R/\eta)^2 \bigr], 
\label{3.10}
\end{eqnarray}
where $R \equiv |\bm{x}-\bm{x'}|$. The complete reduced Green's
function is then given by Eq.~(\ref{3.3}), and $G(x,x')$ can be
obtained from Eq.~(\ref{2.12}). 

The calculation for $\alpha < 0$ returns a different expression for  
$V$, but this can be obtained directly from Eq.~(\ref{3.10}) by
performing the exchange $\eta \leftrightarrow \eta'$. These 
expressions are in fact identical, because the two-point
function satisfies the reciprocity relation 
$V(\eta',\bm{x'};\eta,\bm{x}) = V(\eta,\bm{x};\eta',\bm{x'})$. This
property is derived, for example, in Refs.~\cite{poisson:04b,
poisson:04c}. We conclude that Eq.~(\ref{3.10}) gives the correct
expression for $V$ for both positive and negative values of $\alpha$.     

The Appell function reduces to an ordinary hypergeometric function
when the second argument is zero:
$F_4(\alpha,\beta;\gamma,\gamma';x,0) \equiv
F(\alpha,\beta;\gamma;x)$. This implies that Eq.~(\ref{3.10})
simplifies to 
\begin{eqnarray} 
V(\eta,\bm{x};\eta',\bm{x}) &=& \frac{(2l-1)!!}{(2l-2)!!}
\frac{(\eta/\eta')^l}{\eta^2} 
\nonumber \\ & & \hspace*{-40pt} \mbox{} \times 
F\bigl[-l+1,{\textstyle \frac{3}{2}};-l+{\textstyle \frac{1}{2}};
(\eta'/\eta)^2\bigr] 
\label{3.11}
\end{eqnarray}
when the spatial points $\bm{x}$ and $\bm{x'}$ coincide. When the
spacetime events $x$ and $x'$ fully coincide we have
\begin{equation} 
V(\eta,\bm{x};\eta,\bm{x}) = \frac{l(l+1)}{2\eta^2}, 
\label{3.12}
\end{equation} 
which is recovered after using identity (15.1.20) of
Ref.~\cite{abramowitz-stegun:72}. This result, when combined with
Eqs.~(\ref{2.12}) and (\ref{2.2}), implies that the full retarded
Green's function has the coincidence limit 
\begin{equation} 
G(\eta,\bm{x};\eta-\epsilon,\bm{x}) = \frac{\alpha(\alpha-1)}{2 C^2}
\frac{1}{\eta^{2\alpha + 2}} = \frac{1}{12} {\cal R}, 
\label{3.13}
\end{equation} 
where $\cal R$ is the spacetime's Ricci tensor at $x =
(\eta,\bm{x})$. The presence of $\epsilon = 0^+$ in Eq.~(\ref{3.13})
serves to remind us that the coincidence limit excludes the 
singular, light-cone contribution to the Green's function; this
$\epsilon$ plays the same role in Eqs.~(\ref{2.9}) and (\ref{2.10}).  

The general expression of Eq.~(\ref{3.10}) reduces to simple forms
when $l$ is a small integer. We list a few in Table I. The result for
$l=1$ was first derived in paper I \cite{burko-etal:02}. The general
expression of Eq.~(\ref{3.10}), and the additional special cases
listed in the Table, are new. 

\begin{table*}
\caption{Reduced Green's function $V(\eta,\bm{x};\eta',\bm{x'})$ for
selected values of $l$, expressed in terms of $R =
|\bm{x}-\bm{x'}|$. These expressions are obtained from
Eq.~(\ref{3.10}).}   
\begin{ruledtabular}
\begin{tabular}{ccl} 
$l$ & $\alpha$ & $V = $ \\ 
\hline
$0$ & $\{0,1\}$ & 0 \\ 
$1$ & $\{-1,2\}$ & $(\eta \eta')^{-1}$ \\ 
$2$ & $\{-2,3\}$ & $\frac{3}{2} (\eta^2 + \eta^{\prime 2} 
  - R^2)(\eta \eta')^{-2}$ \\ 
$3$ & $\{-3,4\}$ & $\frac{3}{8} [5\eta^4 
  + 6 \eta^2 \eta^{\prime 2} + 5 \eta^{\prime 4} 
  - 10(\eta^2 + \eta^{\prime 2}) R^2 + 5 R^4](\eta \eta')^{-3}$ \\  
$4$ & $\{-4,5\}$ & $\frac{5}{16} (\eta^2 + \eta^{\prime 2} - R^2)  
  [ 7\eta^4 + 2 \eta^2 \eta^{\prime 2} + 7 \eta^{\prime 4} 
  - 14(\eta^2 + \eta^{\prime 2}) R^2 + 7 R^4 ] (\eta \eta')^{-4}$  
\end{tabular} 
\end{ruledtabular} 
\end{table*} 

\section{Mass change of a charge at rest}   
\label{iv}

In this section we calculate the time-changing inertial mass of a
scalar charge that is comoving with the cosmological fluid. Such a
particle moves on a geodesic of the spacetime, and as we shall see in
Sec.~\ref{v}, the right-hand side of Eq.~(\ref{2.9}) vanishes in
this case. The right-hand side of Eq.~(\ref{2.10}) does not, however,
and we shall find that the particle's mass changes with conformal time
$\eta$. Without loss of generality we place the scalar charge at the
spatial position $\bm{x} = \bm{0}$. 

We begin by deriving a convenient expression for the mass function. 
Equation (\ref{2.10}) can be written as 
\[
\frac{dm}{d\tau} = -q^2 \biggl[ \frac{1}{12} {\cal R} 
+ \int_{0}^{\tau-\epsilon} \frac{d}{d\tau} 
G\bigl(\tau,\tau'\bigr)\, d\tau' \biggr], 
\]
or as 
\[
\frac{dm}{d\tau} = -q^2 \biggl[ \frac{1}{12} {\cal R}  
- G(\tau,\tau-\epsilon) \biggr] 
- q^2 \frac{d}{d\tau} \int_{0}^{\tau-\epsilon}  
G\bigl(\tau,\tau'\bigr)\, d\tau'.  
\]
Taking into account Eq.~(\ref{3.13}) we find that the first term
vanishes, and that the equation can be directly integrated. This gives 
\begin{equation} 
m(\tau) = m(\tau_0) - q^2 \int_{\tau_0}^{\tau-\epsilon}  
G\bigl(\tau,\tau'\bigr)\, d\tau'. 
\label{4.1}
\end{equation} 
We have truncated the domain of integration from $\tau' = 0$ to
$\tau' = \tau_0 > 0$ to avoid problems associated with the big bang 
singularity; in effect we are assuming that the scalar charge was
created at $\tau = \tau_0$ with an initial mass $m(\tau_0)$. As a
final step we change the variable of integration from $\tau'$ to
$\eta'$, using the differential relation $d\tau' = a(\eta')\, d\eta'$
appropriate for a particle at rest, and we re-introduce the reduced
Green's function of Eq.~(\ref{2.12}). Taking into account
Eq.~(\ref{3.3}) we arrive at a mass function given by 
\begin{equation} 
m(\eta) = m_0 - \frac{q^2}{a(\eta)} \int_{\eta_0}^\eta
V(\eta,\bm{0};\eta',\bm{0})\, d\eta', 
\label{4.2}
\end{equation}
where $m_0 \equiv m(\eta_0)$ is the particle's mass at the moment of 
creation. 

A more concrete expression for $m(\eta)$ is obtained by substituting 
Eq.~(\ref{3.11}) into Eq.~(\ref{4.2}). The resulting expression
simplifies when we change the variable of integration from $\eta'$ to 
$\mu \equiv (\eta'/\eta)^2$, which we recognize as the argument of the
hypergeometric function. Simple manipulations yield 
\begin{eqnarray} 
\frac{m(\eta)}{m_0} &=& 1 - \frac{(2l-1)!!}{(2l-2)!!} \zeta  
(\eta_0/\eta)^{\alpha+1} \mbox{sign}(\alpha) 
\nonumber \\ & & \hspace*{-50pt} \mbox{} \times 
\int_{(\eta_0/\eta)^2}^1 \mu^{-(l+1)/2} 
F\bigl(-l+1,{\textstyle \frac{3}{2}};-l+{\textstyle \frac{1}{2}};
\mu \bigr)\, d\mu, 
\label{4.3}
\end{eqnarray} 
where 
\begin{equation}
\zeta \equiv \frac{q^2}{2 m_0 C |\eta_0|^{\alpha+1}} 
\label{4.4}
\end{equation} 
is a dimensionless parameter that determines the strength of the
scalar field's contribution to the mass. We recall from
Eqs.~(\ref{3.1}) and (\ref{3.2}) that $\alpha = -l$ when $\alpha$ is
negative, and $\alpha = l + 1$ when $\alpha$ is positive. We
recall also that when $\alpha < 0$, both $\eta$ and $\eta_0$ are
negative, which implies that $(\eta_0/\eta)^2 \geq 1$; when 
$\alpha > 0$ we have instead $(\eta_0/\eta)^2 \leq 1$.   

Because the hypergeometric function that appears in Eq.~(\ref{4.3}) is
a simple polynomial of degree $l-1$, closed-form expressions for the
mass function can easily be obtained for small values of $\alpha$. We
list a few examples in Table II, and Figs.~1 and 2 provide graphs of 
the selected mass functions. 

\begin{table}
\caption{Mass function for selected values of $\alpha$, expressed in
terms of $t \equiv \eta/\eta_0$. When $\alpha > 0$ $t$ increases from
1 to $\infty$. When $\alpha < 0$ $t$ decreases from 1 to 0. These
expressions are obtained from Eq.~(\ref{4.3}).}  
\begin{ruledtabular}
\begin{tabular}{rl} 
$\alpha$ & $m(\eta)/m_0 = 1 - \zeta (\cdots) $ \\ 
\hline
$2$  & $2\ln(t) t^{-3}$ \\ 
$3$  & $3 t^{-3} - 3 t^{-5}$ \\
$4$  & $\frac{15}{8} t^{-3} + \frac{9}{2} \ln(t) t^{-5} 
          - \frac{15}{8} t^{-7}$ \\ 
$5$  & $\frac{35}{24} t^{-3} + \frac{45}{8} t^{-5} 
          - \frac{45}{8} t^{-7} - \frac{35}{24} t^{-9}$ \\
\hline
$-1$ & $-2\ln(t)$ \\ 
$-2$ & $3 - 3 t^2$ \\ 
$-3$ & $\frac{15}{8} - \frac{9}{2} \ln(t) t^2 - \frac{15}{8} t^4$ \\ 
$-4$ & $\frac{35}{24} + \frac{45}{8} t^2 - \frac{45}{8} t^4 
        - \frac{35}{24} t^6$
\end{tabular} 
\end{ruledtabular} 
\end{table} 
 
\begin{figure}
\includegraphics[angle=-90,scale=0.33]{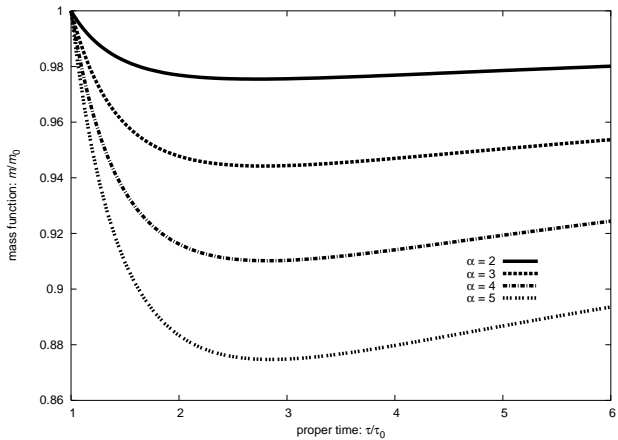}
\caption{Mass function for selected positive values of $\alpha$,
plotted as a function of proper time $\tau/\tau_0 =
(\eta/\eta_0)^{\alpha+1}$. The mass initially decreases but it returns
to its initial value as $\tau \to \infty$. The graph was obtained from
Eq.~(\ref{4.3}) by setting $\zeta = 0.1$.}    
\end{figure}

\begin{figure}
\includegraphics[angle=-90,scale=0.33]{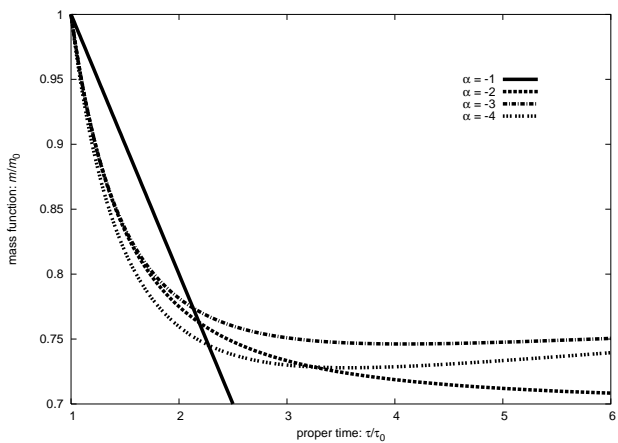}
\caption{Mass function for selected negative values of $\alpha$,
plotted as a function of proper time $\tau/\tau_0$. When $\alpha \neq
-1$, $\tau/\tau_0 = (\eta/\eta_0)^{\alpha+1}$, and the mass initially
decreases but returns to a fraction of its initial value as $\tau \to
\infty$. When $\alpha = -1$ (de Sitter spacetime), $\tau/\tau_0 = 1 -
\ln(\eta/\eta_0)$, and the mass decreases monotonically. The graph was
obtained from Eq.~(\ref{4.3}) by setting $\zeta = 0.1$.}     
\end{figure}

The plots of Fig.~1 suggest that when $\alpha$ is a positive integer,
the mass function decreases at early times, but then increases until
it is restored to its original value. Indeed, it is not difficult to
show that the property $m(\eta)/m_0 \to 1$ as $\eta/\eta_0 \to
\infty$ follows directly from Eq.~(\ref{4.3}) when $\alpha > 0$. The
plots of Fig.~2, on the other hand, suggest that when $\alpha \leq
-2$, only a fraction of the initial mass is eventually returned to the 
particle. This is confirmed by a similar calculation, which reveals 
that when $\alpha < 0$, 
\begin{equation} 
\frac{m(\eta)}{m_0} \to 1 - \frac{2}{l-1} \frac{(2l-1)!!}{(2l-2)!!}
\zeta 
\label{4.5}
\end{equation}
as $\eta/\eta_0 \to 0$ (recall that $l = -\alpha$). When $\alpha = -1$
(de Sitter spacetime) the mass steadily decreases, and the limit goes
to $-\infty$.  

The classical theory of point scalar charges allows the particle's
inertial mass to become negative. This may happen when the parameters
$\alpha$, $q$, $m_0$, and $\eta_0$ are suitably set ($|\alpha|$ or 
$\zeta$ must be sufficiently large). A more complete theory would
provide a description of what happens when $m$ goes to zero. One
possibility is that the particle and its scalar charge simply cease to
exist when the mass is completely radiated away; this scenario is
possible because the theory does not demand scalar-charge
conservation. Another possibility is that the particle continues to
exist as a massless charged object \cite{kazinski-sharapov:03}. We
shall remain neutral on the issue of continuing the life of the scalar
charge beyond $m = 0$.   

\section{Self-force on a slowly moving scalar charge} 
\label{v} 

In this section we allow the scalar charge to move slowly relative to
the cosmological fluid, we calculate the self-force that is acting on
the particle, and we integrate the equations of motion to determine
the particle's world line. To keep technicalities at a minimum we
adopt a matter-dominated cosmology and choose $\alpha = 2$, and we
impose the condition $|\bm{v}| \ll 1$, where $\bm{v} = d\bm{z}/d\eta$
is the particle's coordinate velocity vector. The metric of the
spacetime is given by Eq.~(\ref{2.1}) with $a(\eta) = C \eta^2$. 

The slow-motion assumption implies that the particle's velocity vector 
can be approximated by $u^\mu = (1,\bm{v})/a(\eta)$. The spatial
components of the covariant acceleration are then  
\begin{equation} 
\frac{D u^a}{d\tau} = \frac{1}{a^2} \biggl( \frac{d v^a}{d\eta} +
\frac{2}{\eta} v^a \biggr). 
\label{5.1}
\end{equation}  
The velocity of a particle moving on a geodesic therefore satisfies 
$d\bm{v}/d\eta = -2\bm{v}/\eta$, and in time the particle is  
driven toward the same state of motion ($\bm{v} = \bm{0}$) as the
cosmological fluid. From a Newtonian point of view we might say that
the cosmological expansion exerts a dragging force on the particle
which compels it to come to an eventual rest. We shall adopt this
point of view and refer to the $2\bm{v}/\eta$ term in the equations of
motion as the ``cosmological dragging force''.    

The self-force acting on a scalar charge can be decomposed into
``Ricci'' and ``tail'' forces. The Ricci force appears as the first
term on the right-hand side of Eq.~(\ref{2.9}), while the tail force
involves an integral over the past portion of the world line. For a
matter-dominated cosmology the Ricci tensor is given by 
$R^x_{\ x} = R^y_{\ y} = R^z_{\ z} = -R^\eta_{\ \eta} 
= 6/(C^2 \eta^6)$, and the spatial components of the Ricci force are  
\begin{equation} 
F^a_{\rm Ricci} = \frac{1}{a^2} \frac{2 q^2}{C \eta^4} v^a. 
\label{5.2}
\end{equation} 
Notice that the Ricci force is directed in the direction of $\bm{v}$. 
For a matter-dominated cosmology the retarded Green's function is
given by $G = V/[a(\eta)a(\eta')]$ with $V = 1/(\eta\eta')$, so that
$G = C^{-2} (\eta\eta')^{-3}$. Evaluation of the tail integral on the
right-hand side of Eq.~(\ref{2.9}) gives 
\begin{equation} 
F^a_{\rm tail} = \frac{1}{a^2} \biggl( - \frac{3 q^2}{C \eta^4} \ln
\frac{\eta}{\eta_0} \biggr) v^a 
\label{5.3}
\end{equation} 
for the tail force. Notice that we have again truncated the tail
integral so that it begins at $\eta = \eta_0$. Notice also that the
tail force is directed in the direction opposite to $\bm{v}$.  

Combining Eqs.~(\ref{5.1})--(\ref{5.3}) gives us the particle's
equations of motion, 
\begin{equation} 
\frac{d \bm{v}}{d \eta} = - \frac{2}{\eta} \bm{v}  
+ \frac{q^2}{C \eta^4 m(\eta)} \biggl( 2 - 3 \ln \frac{\eta}{\eta_0} 
\biggr) \bm{v}. 
\label{5.4}
\end{equation} 
The first term on the right-hand side of Eq.~(\ref{5.4}) is the
cosmological dragging force, which is directed along $-\bm{v}$. The
second term is the Ricci force, and this is directed along
$\bm{v}$. Finally, the third term is the tail force, and this is
directed along $-\bm{v}$. The sign of the self-force depends
on time: The Ricci force dominates at early times ($\eta/\eta_0$ close
to unity) and it pushes on the particle, while the tail force
dominates at late times and pulls on the particle. 

We shall assume that $q^2$ is small and that the self-force produces
only a small correction to the particle's motion. In this regime we
can replace $m(\eta)$ by $m_0$ on the right-hand side of
Eq.~(\ref{5.4}), and our precise requirement is that 
\begin{equation} 
\zeta \equiv \frac{q^2}{2 m_0 C \eta_0^3} \ll 1; 
\label{5.5}
\end{equation}  
notice that unlike $q^2$, $\zeta$ is dimensionless. Introducing $t
\equiv \eta/\eta_0 \geq 1$ we rewrite Eq.~(\ref{5.4}) as
\begin{equation} 
\frac{d \bm{v}}{dt} = - \frac{2}{t} \bm{v}  
+ \frac{2 \zeta}{t^4} \Bigl( 2 - 3 \ln t \Bigr) \bm{v}. 
\label{5.6}
\end{equation} 
The largest term on the right of Eq.~(\ref{5.6}) is the cosmological
dragging force, and this is only slightly offset by the self-force. 
Integration of Eq.~(\ref{5.6}), in the regime $\zeta \ll 1$, produces 
\begin{equation} 
\bm{v}(t) = \frac{\bm{v}(1)}{t^2} \biggl[ 1 + \frac{2\zeta}{3t^3} 
\Bigl( t^3 - 1 + 3\ln t \Bigr) + O(\zeta^2) \biggr].  
\label{5.7}
\end{equation} 
This result shows that when a charged ($\zeta \neq 0$) particle and
an uncharged ($\zeta = 0$) particle start with the same initial
velocity $\bm{v}(1)$, the charged particle moves faster at times
$t > 1$. Ultimately, however, both particles are brought to rest
by the cosmological expansion.    

\section{Self-force on a slowly moving electric charge} 
\label{vi} 

That a charged particle moves faster than an uncharged particle in a
cosmological spacetime (starting with equal initial velocities) is
not an artifact of the scalar theory. The same conclusion applies in
the context of electrodynamics. In this section we consider a point
particle of (constant) mass $m$ and electric charge $e$ that moves
slowly relative to the cosmological fluid. We once more calculate the
self-force and integrate the particle's equations of motion. Here we
are able to select a more general class of cosmological models, and we 
let the scale factor take its general form $a(\eta) = C \eta^\alpha$
with $\alpha$ no longer restricted to integers; we do, however,
restrict $\alpha$ to be positive in order to keep the discussion
simple. We once more impose the condition $|\bm{v}| \ll 1$.  

It is known that the curved-spacetime version of Maxwell's theory is 
conformally invariant (see, for example, Appendix D of
Ref.~\cite{wald:84}), and the metric of Eq.~(\ref{2.1}) is clearly
conformally related to flat spacetime. This implies that the
electromagnetic field produced by the moving charge propagates
without tails, and that the self-force is purely local. The
charge's equations of motion were first determined by Hobbs
\cite{hobbs:68}, who corrected the earlier pioneering treatment of 
DeWitt and Brehme \cite{dewitt-brehme:60} (see also the reviews
\cite{poisson:04b} and \cite{poisson:04c} for self-contained
derivations). In a conformally-flat spacetime, and in the absence of
an external force, the particle moves according to 
\begin{equation} 
m \frac{D u^\mu}{d\tau} = \frac{1}{3} e^2 \bigl( \delta^\mu_{\ \nu}  
+ u^\mu u_\nu \bigr) R^\nu_{\ \lambda} u^\lambda; 
\label{6.1}
\end{equation} 
here the self-force contains only a Ricci term, and the tail force is 
absent. 

The nonvanishing components of the Ricci tensor are $R^\eta_{\ \eta} =
-3\alpha C^{-2} \eta^{-2\alpha-2}$ and $R^x_{\ x} = R^y_{\ y} 
= R^z_{\ z} = \alpha(2\alpha-1) C^{-2} \eta^{-2\alpha-2}$. In the
slow-motion approximation the particle's velocity vector is
$u^\mu = (1,\bm{v})/a(\eta)$, where $\bm{v} = d\bm{z}/d\eta$ is the
coordinate velocity. Using this information, we find that the spatial
components of Eq.~(\ref{6.1}) reduce to 
\begin{equation} 
\frac{d\bm{v}}{d\eta} = -\frac{\alpha}{\eta} \bm{v} + \frac{2}{3}
\alpha(\alpha+1) \frac{e^2}{m C} \frac{1}{\eta^{\alpha+2}} \bm{v}. 
\label{6.2}
\end{equation} 
The first term on the right-hand side of Eq.~(\ref{6.2}) is the
cosmological dragging force; it is directed along $-\bm{v}$. The
second term is the electromagnetic self-force; it is directed along
$\bm{v}$ and it pushes on the particle.  

In order to integrate Eq.~(\ref{6.2}) we rewrite it as 
\begin{equation}
\frac{d\bm{v}}{d t} = -\frac{\alpha}{t} \bm{v} + \frac{4}{3}
\alpha(\alpha+1) \zeta \frac{1}{t^{\alpha+2}} \bm{v}, 
\label{6.3}
\end{equation} 
re-introducing $t \equiv \eta/\eta_0$ and defining the small
dimensionless parameter  
\begin{equation} 
\zeta = \frac{e^2}{2 m C \eta_0^{\alpha+1}}. 
\label{6.4}
\end{equation} 
Integration of Eq.~(\ref{6.3}) for $\zeta \ll 1$ is straightforward,
and we obtain 
\begin{equation} 
\bm{v}(t) = \frac{\bm{v}(1)}{t^\alpha} \biggl[ 1 
+ \frac{4\alpha \zeta}{3 t^{\alpha+1}} \Bigl( t^{\alpha+1}-1 \Bigr)  
+ O(\zeta^2) \biggr]. 
\label{6.5}
\end{equation} 
This result leads to the same conclusion as in Sec.~\ref{v}: When a
charged ($\zeta \neq 0$) particle and an uncharged ($\zeta = 0$)
particle start with the same initial velocity $\bm{v}(1)$, the charged
particle moves faster at times $t > 1$. Both particles are
eventually brought to rest by the cosmological expansion.  

\section{Discussion: a pushing self-force?} 
\label{vii} 

It is somewhat surprising that a self-force (scalar or 
electromagnetic) would act so as to push the particle; this goes
against a naive expectation that the self-force should always give
rise to radiative damping. The examples worked out in the preceding
sections illustrate that this expectation is indeed naive. But there
is no need to resort to curved-spacetime examples to show this. The
expectation is naive also in the simplest context of slow-motion
electrodynamics in flat spacetime.  

Consider a particle of mass $m$ and electric charge $e$ moving in the 
field of a charge $-Q$; we take $e$ and $Q$ to be positive quantities,
and we suppose that $Q$ is held fixed at the origin of a Cartesian
coordinate system. The external force acting on $e$ is the Coulomb
force exerted by $Q$: 
\begin{equation} 
\bm{F}_{\rm ext} = -k \frac{eQ}{r^2} \bm{\hat{r}}, 
\label{7.1} 
\end{equation} 
where $k = (4\pi\varepsilon_0)^{-1}$, $r^2 = x^2 + y^2 + z^2$ is the
squared distance from $Q$, and $\bm{\hat{r}} = \bm{x}/r$ is a radial
unit vector; the force of Eq.~(\ref{7.1}) is attractive. The
self-force acting on $e$ is
\begin{equation}  
\bm{F}_{\rm self} = \frac{\mu_0}{6\pi c} \frac{e^2}{m}
\bm{\dot{F}}_{\rm ext}, 
\label{7.2}
\end{equation} 
where an overdot indicates differentiation with respect to time. This 
expression for the self-force is derived, for example, in Sec.~75 of
the Landau and Lifshitz textbook \cite{landau-lifshitz:b2}. It is
usually expressed in terms of the rate of change of the particle's
acceleration vector, but Landau and Lifshitz explain that it is
consistent to make the replacement $\bm{a} \to \bm{F}_{\rm ext}/m$ on
the right-hand side of Eq.~(\ref{7.2}). A first-principles derivation
of this equation is also presented in Ref.~\cite{poisson:99}.  

From Eq.~(\ref{7.1}) we calculate that the rate of change of the
external force is given by $\bm{\dot{F}}_{\rm ext} = -k eQ (\bm{v} - 3
\dot{r} \bm{\hat{r}})/r^3$, where $\dot{r} \equiv \bm{v} \cdot
\bm{\hat{r}}$ is the radial component of the velocity vector. We
decompose this vector in terms of radial and tangential components,   
\begin{equation} 
\bm{v} = \bm{v}_\bot + \dot{r} \bm{\hat{r}}, \qquad
\bm{v}_\bot \cdot \bm{\hat{r}} = 0,
\label{7.3}
\end{equation} 
and we substitute this result into Eq.~(\ref{7.2}). This gives  
\begin{equation} 
\bm{F}_{\rm self} = \frac{\mu_0 k}{6\pi c} \frac{e^3 Q}{m} 
\bigl( -\bm{v}_\bot + 2 \dot{r} \bm{\hat{r}} \bigr) 
\label{7.4}
\end{equation}
for the self-force acting on the moving charge. 

Suppose that the charge $e$ is moving tangentially, so that $\dot{r} =
0$. Then the self-force is directed along $-\bm{v}_\bot = -\bm{v}$ and
we find that it is indeed a dragging force. Now suppose that the
charge is moving radially inward, so that $\bm{v}_\bot = \bm{0}$
and $\dot{r} < 0$. Then the self-force is directed along
$-\bm{\hat{r}}$, that is, in the same direction as the external
force. In this situation the self-force pushes on the particle, in
violation of the naive expectation. 

More generally, the rate of work done by the self-force, 
$\bm{F}_{\rm self} \cdot \bm{v}$, is proportional to 
$-v_\bot^2 + 2 \dot{r}^2$, where $v_\bot \equiv |\bm{v}_\bot|$. When  
$v_\bot > \sqrt{2} \dot{r}$ the work done on the particle is negative
and the self-force exerts a drag. When, however, $v_\bot < \sqrt{2}
\dot{r}$ the work done is positive and the self-force pushes.  

Does a pushing self-force violate energy conservation? The answer is
in the negative, in spite of appearances to the contrary. Quite
generally, the rate of work done on a particle of charge $e$ by the 
self-force (which we now express in terms of the particle's
acceleration $\bm{a}$) is 
\[
\bm{F}_{\rm self} \cdot \bm{v} = \frac{\mu_0 e^2}{6\pi c} \bm{\dot{a}} 
\cdot \bm{v} = \frac{\mu_0 e^2}{6\pi c} \biggl[ \frac{d}{dt} 
(\bm{a} \cdot \bm{v}) - |\bm{a}|^2 \biggr]. 
\]
The last term on the right-hand side gives the electromagnetic power 
radiated by the particle, and we have the statement 
\begin{equation} 
[\mbox{rate of work done}] + [\mbox{radiated power}] = 
\frac{\mu_0 e^2}{6\pi c} \frac{d}{dt} (\bm{a} \cdot \bm{v}). 
\label{7.5}
\end{equation} 
Integrating this equation over a finite time interval gives rise to
the following conservation statement: 
\begin{eqnarray} 
[\mbox{work done}] + [\mbox{radiated energy}] &=& 
\nonumber \\ & &  
\hspace*{-120pt} 
[\mbox{total change in
$\mu_0 e^2 (\bm{a} \cdot \bm{v})/(6\pi c)$}].
\label{7.6}
\end{eqnarray} 
The term on the right can be interpreted as a change in the system's
``internal energy'' as it moves from one configuration to
another. (The system is the charged particle plus the agent
responsible for the external force.) This form of energy is associated
with the part of the electromagnetic field that stays bound to the
system instead of propagating freely as electromagnetic radiation.  

In some circumstances the total change in internal energy is
zero. For example, when the motion is periodic the total change
vanishes after one complete period. As another example, the total
change vanishes when the motion proceeds from a state of no
acceleration to another state of no acceleration. In such
circumstances the statement of Eq.~(\ref{7.6}) simplifies to 
$[\mbox{work done}] + [\mbox{radiated energy}] = 0$, and the work done
by the self-force must be negative. In more general circumstances,
however, the statement of energy conservation must include the change
in internal energy, and it becomes possible for the self-force to do
positive work on the particle. This was well illustrated in the
previous discussion. 

\acknowledgments 

This work was supported by the Natural Sciences and Engineering
Research Council of Canada. 

\bibliography{motion}

\begin{thebibliography}{10}
\expandafter\ifx\csname url\endcsname\relax
  \def\url#1{{\tt #1}}\fi
\expandafter\ifx\csname urlprefix\endcsname\relax\def\urlprefix{URL }\fi

\bibitem{LISA}
The LISA web site is located at http://lisa.jpl.nasa.gov/.

\bibitem{poisson:04b}
E.~Poisson, {\em The motion of point particles in curved spacetime\/}, Living
  Rev. Relativity {\bf 7} (2004), 6. [Online article]: cited on \today,
  http://www.livingreviews.org/lrr-2004-6.

\bibitem{poisson:04c}
E.~Poisson, {\em Radiation reaction of point particles in curved spacetime\/},
  Class. Quantum Grav. {\bf 21}, R153 (2004).

\bibitem{poisson:04d}
E.~Poisson, {\em The gravitational self-force\/} (2004),
  http://xxx.lanl.gov/abs/gr-qc/0410127.

\bibitem{burko-etal:02}
L.~M. Burko, A.~I. Harte, and E.~Poisson, {\em Mass loss by a scalar charge in
  an expanding universe\/}, Phys. Rev. D {\bf 65}, 124006 (2002).

\bibitem{quinn:00}
T.~C. Quinn, {\em Axiomatic approach to radiation reaction of scalar point
  particles in curved spacetime\/}, Phys. Rev. D {\bf 62}, 064029 (2000).

\bibitem{wald:84}
R.~W. Wald, {\em General Relativity\/} (Chicago University Press, Chicago,
  1984).

\bibitem{poisson:b04}
E.~Poisson, {\em A relativist's toolkit: The mathematics of black-hole
  mechanics\/} (Cambridge University Press, Cambridge, England, 2004).

\bibitem{abramowitz-stegun:72}
M.~Abramowitz and I.~A. Stegun, {\em Handbook of Mathematical Functions\/}
  (Dover, New York, 1972).

\bibitem{gradshteyn-ryzhik:80}
I.~S. Gradshteyn and I.~M. Ryzhik, {\em Table of Integrals, Series, and
  Products\/} (Academic Press, Orlando, 1980).

\bibitem{kazinski-sharapov:03}
P.~Kazinski and A.~Sharapov, {\em Radiation reaction for a massless charged
  particle\/}, Class. Quantum Grav. {\bf 20}, 2715 (2003).

\bibitem{hobbs:68}
J.~M. Hobbs, {\em A veirbein formalism for radiation damping\/}, Ann. Phys.
  (N.Y.) {\bf 47}, 141 (1968).

\bibitem{dewitt-brehme:60}
B.~S. DeWitt and R.~W. Brehme, {\em Radiation damping in a gravitational
  field\/}, Ann. Phys. (N.Y.) {\bf 9}, 220 (1960).

\bibitem{landau-lifshitz:b2}
L.~D. Landau and E.~M. Lifshitz, {\em The Classical Theory of Fields, Fourth
  Edition\/} (Butterworth-Heinemann, Oxford, England, 2000).

\bibitem{poisson:99}
E.~Poisson, {\em An introduction to the Lorentz-Dirac equation\/} (1999),
  http://xxx.lanl.gov/abs/gr-qc/9912045.

\end{thebibliography}
\end{document}